\documentclass[]{interact}

\usepackage[dvipsnames]{xcolor}
\usepackage{pgfplots}	
\graphicspath{{./fig/}}
\pgfplotsset{compat=1.18}

\usepackage{epstopdf}
\usepackage[caption=false]{subfig}
\usepackage{mathrsfs}

\usepackage[numbers,sort&compress]{natbib}
\bibpunct[, ]{[}{]}{,}{n}{,}{,}
\renewcommand\bibfont{\fontsize{10}{12}\selectfont}

\theoremstyle{plain}
\newtheorem{theorem}{Theorem}[section]
\newtheorem{lemma}[theorem]{Lemma}

\newtheorem{proposition}[theorem]{Proposition}

\theoremstyle{definition}

\theoremstyle{remark}
\newtheorem{remark}{Remark}

\newcommand{\rot}[1]{{\color{red}#1}}
\newcommand{\blau}[1]{{\color{blue}#1}}

\newcommand{\Div}{\operatorname{Div}}
\newcommand{\Grad}{\operatorname{Grad}}
\newcommand{\Sym}{{\operatorname{sym}}}
\newcommand{\trace}{{\operatorname{tr}}}

\def\ttM{\mathrm{M}}
\def\ttD{\mathrm{D}}
\def\ttB{\mathrm{B}}
\def\ttG{\mathrm{G}}
\def\ttN{\mathrm{N}}
\def\ttv{\mathrm{v}}
\def\ttu{\mathrm{u}}
\def\ttS{\mathrm{S}}
\def\ttlam{\mathrm{\lambda}}
\def\tttau{\mathrm{\tau}}
\def\ttb{\mathrm{b}}
\def\ttf{\mathrm{f}}
\def\ttg{\mathrm{g}}
\def\tth{\mathrm{h}}
\def\tty{\mathrm{y}}
\def\ttz{\mathrm{z}}
\def\Dtau{\tau}
\def\dtau{d_\tau}
\def\T{\mathcal{T}}
\def\RR{\mathbb{R}}

\begin{document}

\articletype{ARTICLE}

\title{On the velocity-stress formulation for geometrically nonlinear elastodynamics and its structure-preserving discretization}

\author{
\name{Tobias Thoma\textsuperscript{a}\thanks{CONTACT T. Thoma. Email: tobias.thoma@tum.de}, Paul Kotyczka\textsuperscript{a} and Herbert Egger\textsuperscript{b}}
\affil{\textsuperscript{a}Technical University of Munich, Germany; TUM School of Engineering
	and Design, Chair of Automatic Control}
\textsuperscript{b}Johannes Kepler University Linz, Austria; Institute of Numerical Mathematics
}

\maketitle

\begin{abstract}
We consider the dynamics of an elastic continuum under large deformation but small strain. Such systems can be described by the equations of  geometrically nonlinear elastodynamics in combination with the St.~Venant-Kirchhoff material law. 
The velocity-stress formulation of the problem turns out to have a formal port-Hamiltonian structure. In contrast to the linear case, the operators of the problem are modulated by the displacement field which can be handled as a passive variable and integrated along with the velocities. 
A weak formulation of the problem is derived and essential boundary conditions are incorporated via Lagrange multipliers. 
This variational formulation explicitly encodes the transfer between kinetic and potential energy in the interior as well as across the boundary, thus leading to a global power balance and ensuring passivity of the system.  
The particular geometric structure of the weak formulation can be preserved under Galerkin approximation via appropriate mixed finite elements. 
In addition, a fully discrete power balance can be obtained by appropriate time discretization. 
The main properties of the system and its discretization are shown theoretically and demonstrated by numerical tests.
\end{abstract}

\begin{keywords}
nonlinear elastodynamics;  velocity-stress formulation; port-Hamiltonian systems; structure-preserving discretization; mixed finite elements; 
\end{keywords}

\def\herbert#1{{\color{blue}}#1}

\section{Introduction}
\label{sec:introduction}
Linear models of elastodynamics are extensively used in structural analysis and engineering design and efficient finite element methods have been developed for simulation and control. 
In~\cite{Makridakis1992} mixed finite elements for the \textit{displacement-stress} and \textit{velocity-stress} formulation of linear elastodynamics were investigated, partly based on corresponding results for the wave equation~\cite{Geveci1988}.
A new family of mixed finite elements for linear elastodynamics suitable for mass lumping was presented in \cite{Becache2002}.
An overview about the construction and analysis of variational numerical schemes for time-dependent wave propagation problems is given in \cite{Cohen2002,Joly2003}. 
Primal-dual and dual-primal velocity-stress formulations for linear elastodynamics are considered as examples, and stability and energy conservation are derived for semi- and fully-discrete schemes. 
The velocity-stress formulation is also frequently used for simulations in seismology~\cite{Festa2005}.
In~\cite{Arnold2014}, dual mixed finite element methods for elastodynamics with weakly imposed symmetry for the stresses have been investigated. %
Another efficient stabilized nodal finite element method for elastodynamics was proposed in \cite{Scovazzi2017}. Their framework is based on a space-time variational statement of the velocity-stress formulation and is compatible with weakly and strongly enforced Dirichlet boundary conditions.
In \cite{Zhang2019}, a dual velocity-stress formulation with weakly imposed symmetry was analyzed in the context of the virtual finite element approach. 
A mixed finite element method based on a \textit{port-Hamiltonian} (PH) formulation was used in \cite{Brugnoli2021a} to simulate, in a power-preserving way, a coupled system of linear thermoelasticity. 

The PH framework provides a general approach for modeling, analysis, and control of complex multi-physical dynamical systems~\cite{Duindam2009}. 
Originating in a synthesis of network and Hamiltonian perspectives for \emph{open}, i.e., controlled, finite-dimensional systems \cite{maschke1992port}, the past two decades brought important progress in the development of the theory for infinite-dimensional systems from different physical domains, including structural mechanics; see e.g. \cite{schaft2002hamiltonian,legorrec2005dirac,Macchelli2009,jacob2012linear} and the review article \cite{Rashad2020}.
In this context, also the structure-preserving numerical approximation has been a major field of interest; see \cite{golo2004hamiltonian,farle2013port,Kotyczka2018WeakFO,Cardoso-Ribeiro2020} and, in particular, \cite{Brugnoli2019a,Brugnoli2019b,Brugnoli2020,Warsewa2021,Brugnoli2021c} for applications in linear elastodynamics. 
Recent contributions to the modeling and structure-preserving discretization of nonlinear structures have been presented in~\cite{Brugnoli2021a,Brugnoli2021b,thoma2022,Kinon2023}.
For discretized and finite-dimensional systems, symplectic integration~\cite{kotyczka2019discreteSCL}, discrete gradients~\cite{falaize2016passive}, Petrov-Galerkin~\cite{egger2021energy}, and projection methods~\cite{mueller2021time} 
have been proposed for the structure-preserving time discretization.

In this paper, we consider the motion of an elastic continuum described by the equations of geometrically nonlinear elastodynamics. The corresponding  velocity-stress formulation has a PH structure which, in contrast to linearized models, is modulated by the displacement field. 
The main structural characteristics of the problem become apparent from a weak formulation in which Dirichlet boundary conditions are incorporated by Lagrange multipliers.
A direct consequence of the variational formulation of the equations is a power balance which also ensures passivity of the system.
In contrast to related work, we here advocate a representation which is split into (a) the velocity-stress model with the body forces acting as distributed input and (b) an additional kinematic equation, which is integrated to yield the modulating displacement field. 
This particular representation explicitly encodes the different roles of the system variables with respect to the power flows and it is also very well-suited for systematic discretization.
The Galerkin approximation by mixed finite elements in space automatically preserves the geometric structure and leads to a finite-dimensional PH system in form of a differential-algebraic equation (DAE). The subsequent time discretization by the implicit midpoint-rule yields a corresponding power balance also on the fully discrete level. 
Numerical results are presented for the movement of a robot arm in order to illustrate the theoretical findings.

The remainder of the article is organized as follows: In Section \ref{sec:2}, we recall the basic equations of geometrically nonlinear mechanics and introduce the velocity-stress formulation of our problem. In addition, we derive its weak formulation and the underlying power balance. The structure-preserving discretization by mixed finite elements in space and the implicit midpoint rule in time is discussed in Section~\ref{sec:3}, and  
Section~\ref{sec:4} presents some numerical tests for the proposed methods. A brief discussion of our main results and some concluding remarks are given in Section~\ref{sec:5}.

\section{Modeling}
\label{sec:2}

We start with recalling the governing equations of nonlinear continuum mechanics for a St. Venant–Kirchhoff material. After that, we present the velocity-stress formulation of the system and discuss the underlying modulated PH structure. We then derive an appropriate weak form of the governing equations and establish the inherent power balance for the system.

\subsection{Nonlinear continuum mechanics}

\begin{figure}[htbp]
	\centering
	\def\svgwidth{7.5cm}
	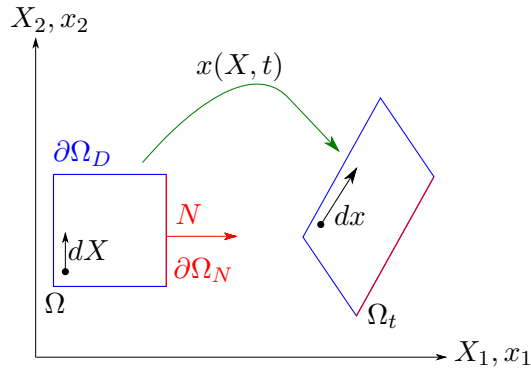
	\caption{Mapping of an infinitesimal line element $dX$ from the material (undeformed) configuration ($\Omega\subset\mathbb{R}^2$) to the spatial (deformed) one ($\Omega_t\subset\mathbb{R}^2$) via the deformation gradient $F = \operatorname{Grad}(x)$, $dx=F\cdot dX$.}
	\label{fig:F}
\end{figure}

Following standard practice~\cite{Bonet2008,Wriggers2001},
we use material coordinates. 
All fields therefore depend on the spatial coordinate $X$ and time $t$, which we mostly omit in our notation. 
The momentum balance of the system is given by 
\begin{equation}
	\label{eq:impulse}
	\rho\ddot{u} = \text{Div}(F\cdot S) + b \qquad \text{in } \Omega,
\end{equation}
where $\Omega \subset\mathbb{R}^d$, $d=1,2,3$, is the reference configuration of the body. 
As usual, $\rho$ denotes the material density, $u(X,t)=x(X,t)-X$ the displacement field, and 
\begin{equation}
    \label{eq:deformation-gradient}
    F=\Grad(u)+I    
\end{equation}
the deformation gradient; see Fig.~\ref{fig:F} for a sketch. 
Furthermore, $S$ is the second Piola-Kirchoff stress tensor, and $b$ the vector of volume forces caused by gravitation or other exogenous loads. 
In this paper, we restrict our considerations to St. Venant-Kirchhoff materials which are characterized by a linear stress-strain relation
\begin{equation}
    \label{eq:Piola-Kirchhoff-2}
	S = C:E.
\end{equation}
The fourth order stiffness tensor $C$ is assumed symmetric and positive definite, for instance $C : E = 2 \mu E + \lambda \trace(E) I$, and it acts on the nonlinear Green strain tensor
\begin{equation}
	\label{eq:state_E}
	E = \frac{1}{2}(F^\top\cdot F - I).
\end{equation}
To complete the system description, we consider mixed boundary conditions
\begin{subequations}
\begin{align}
	u &= u_D \qquad \text{on } \partial\Omega_D, \\
	F\cdot S\cdot N &= \tau_N \qquad \text{on }\partial\Omega_N.
\end{align}
\end{subequations}
Here $\partial\Omega=\partial\Omega_D\cup\Omega_N$ denotes an appropriate decomposition of the boundary into Dirichlet and Neumann part, $N$  denotes the outer oriented normal vector field on $\partial\Omega$, and $u_D$, $\tau_N$ are the given boundary data.
The choice of appropriate initial conditions will be discussed later on.

\subsection{Velocity-stress formulation}
By introducing the velocity ${v=\dot{u}}$ and the
compliance tensor $A = C^{-1}$, the above equations can be rewritten equivalently as a first order system
\begin{subequations}
\begin{align}
	\label{eq:firstorder:a}
	\rho\dot{v} &= \Div(F(u)\cdot S) + b \\
	\label{eq:firstorder:b}
	A:\dot{S} &= \Sym(F(u)^\top\cdot\Grad(v)).
\end{align}
\end{subequations}
Rewriting the Dirichlet boundary conditions leads to
\begin{subequations}
\begin{align}
	\label{eq:bc:a}
	v &= v_D \quad \text{on }\partial\Omega_D, \\
	\label{eq:bc:b}
	F(u)\cdot S\cdot N &= \tau_N \quad \text{on } \partial\Omega_N,
\end{align}
\end{subequations}
where $v_D = \dot u_D$ is the prescribed velocity at the Dirichlet boundary $\partial\Omega_D$. 
The \emph{velocity-stress formulation} \eqref{eq:firstorder:a}--\eqref{eq:firstorder:b} does not provide a complete state representation, but has to be complemented by a kinematic equation
\begin{equation}
    \label{eq:kinematic-eq}
    \dot u = v. \tag{6c}
\end{equation}
This equation allows to infer information about the displacement field $u$, which \emph{modulates} the system dynamics \eqref{eq:firstorder:a}--\eqref{eq:firstorder:b} via the deformation gradient $F(u)$. 

\begin{remark}
	\label{rem:lin}
Substituting $F(u)$ by the identity matrix $I$ leads to the velocity-stress formulation of linear elastodynamics~\cite{Makridakis1992}, which is governed by the linearized strain tensor $\epsilon = \Sym(\Grad(u))$ instead of $E$, and the Cauchy stress $\sigma$ instead of $S$. 
\end{remark}

\subsection{Hamiltonian and power-conjugated variables}

The internal energy of the system \eqref{eq:firstorder:a}--\eqref{eq:kinematic-eq} is given by the Hamiltonian
\begin{equation}
    \label{eq:Hamiltonian-elastic}
    H(p,E) = \int_\Omega \frac{1}{2 \rho} p \cdot p + \frac{1}{2} (C:E):E \, dX,
\end{equation}
consisting of the kinetic and elastic (potential) energy. 
The work of the body forces will be treated as external contribution in our description. 
We choose the momentum density $p = \rho v$ and the Green strain tensor $E$ as the \emph{state variables}, in compliance with a unified energy storage model, underlying the PH modeling paradigm or its graphical representation by generalized bond graphs.
The corresponding \emph{co-state variables} given by the variational derivatives
\begin{equation}
    \label{eq:co-states}
    \delta_p H(p,E) = \frac{p}{\rho} = v, \qquad 
    \delta_E H(p,E) = C:E = S
\end{equation}
of the Hamiltonian here correspond to the velocity and stress, respectively. 

\begin{remark}
\label{remark:phs}
The differential equations \eqref{eq:firstorder:a}--\eqref{eq:firstorder:b} may be written compactly as
\begin{align}
    \label{eq:PH-a}
    \begin{bmatrix} 
        \dot p\\ \dot E
    \end{bmatrix}
    &=
   \begin{bmatrix}
        0 & \text{Div}(F(u)\cdot\times) \\
	\Sym(F(u)^\top\cdot\Grad(\times)) & 0
    \end{bmatrix}
    \begin{bmatrix}
        v\\ S
    \end{bmatrix}
    +
    \begin{bmatrix}
        I\\ 0
    \end{bmatrix}
    b,
\end{align}
according to the formulation as a distributed parameter PH system; see e.g.~\cite{Mehrmann2022,schaft2002hamiltonian}. 
In this language, $x=(p,E)$ and $e = (v,S)$ are called the \emph{state} (or energy) and co-state (or co-energy or \emph{effort}) variables. The time derivatives of the states $f = (-\dot p, -\dot E)$, the \emph{flows}, here with the commonly used sign convention, are power-conjugated to the efforts, such that $\dot H + \langle e, f \rangle_\Omega = 0$, with $\langle \cdot, \cdot \rangle_\Omega$ the appropriate duality product. The PH formulation is completed by the definition of \emph{output} variables, which are power-conjugated to the imposed \emph{inputs}, and which naturally emerge in the power balance equation, after possible integration by parts. In the considered case, we have the \emph{distributed port} $(b, v)$ and the \emph{boundary ports} $(v_D, F(u)\cdot S \cdot N)$, $(\tau_N, v)$ on $\partial \Omega_D$ and $\partial \Omega_N$, respectively.
The differential operator on the right hand side, which here is \emph{modulated} by the displacement $u$, is \emph{formally skew-adjoint} and describes the in-domain exchange between kinetic and elastic energy, and the coupling with the boundary ports. 
Related PH representations for beams, plates, and strings can be found in~\cite{Brugnoli2021a,Brugnoli2021b,thoma2022}.
\end{remark}

\subsection{Variational identities}

As a next step, we derive an appropriate weak form of the equations \eqref{eq:firstorder:a}--\eqref{eq:kinematic-eq} which completely encodes the underlying energy-exchange mechanism and reveals the role of input and output variables. 
For ease of notation, we introduce 
$$
\langle a,b \rangle_M=\int_{M} a \odot b \; dX
$$ 
for the integration of products of scalar, vector, and tensor valued functions over some manifold $M$.
We further use a Lagrange multiplier approach to enforce the Dirichlet boundary conditions \eqref{eq:bc:a}, which leads to the following statement.
\begin{lemma} \label{lem:weak}
Let $(v,S,u)$ be a sufficiently regular solution of \eqref{eq:firstorder:a}--\eqref{eq:kinematic-eq} and \eqref{eq:bc:a}--\eqref{eq:bc:b} with $S=S^\top$, and define the reaction force $\lambda:=F(u)\cdot S\cdot N$ on $\partial\Omega_D$. Then the variational identities
\begin{subequations}
\label{eq:weak}
 \begin{align} 
	\langle\rho\dot{v},\delta v\rangle_{\Omega} &= -\langle S,F(u)^\top\cdot\Grad(\delta v)\rangle_{\Omega} + \langle b,\delta v\rangle_{\Omega} \label{eq:weak:a}\\
    &\qquad \qquad \qquad + \langle\lambda,\delta v\rangle_{\partial\Omega_D} + \langle\tau_N,\delta v\rangle_{\partial\Omega_N} \notag
\\
	\langle A:\dot{S},\delta S\rangle_{\Omega} &= \langle F(u)^\top\cdot\Grad(v),\delta S\rangle_{\Omega} 		\label{eq:weak:b}
\\
	\langle v,\delta\lambda\rangle_{\partial\Omega_D} &= \langle v_D,\delta\lambda\rangle_{\partial\Omega_D} 		\label{eq:weak:c}
\end{align}
\end{subequations}
hold for all regular test functions $(\delta v, \delta S, \delta \lambda)$ with $\delta S=\delta S^\top$, and all $t \ge 0$ of interest. 
\end{lemma}

\begin{proof}
By testing equation~\eqref{eq:firstorder:a} and integration by parts, one can see that
\begin{align*}
  \langle\rho\dot{v},\delta v\rangle_{\Omega} 
&= \langle \Div(F(u)\cdot S), \delta v\rangle_{\Omega} + {\langle b,\delta v\rangle}_{\Omega} \\
&= -\langle S,F(u)^\top\cdot\Grad(\delta v)\rangle_{\Omega} + \langle b,\delta v\rangle_{\Omega} + \langle F(u)\cdot S\cdot N,\delta v\rangle_{\partial\Omega}.
\end{align*}
By substituting $\tau_N$ and $\lambda$ for $F(u)\cdot S\cdot N$ on $\partial\Omega_N$ and $\partial\Omega_D$, respectively, we already arrive at \eqref{eq:weak:a}.
Equation~\eqref{eq:weak:b} follows immediately by testing \eqref{eq:firstorder:b} and exploiting the symmetry of $\delta S$, 
which gives rise to 
\begin{align*}
\langle \Sym( F(u)^\top \cdot \Grad(v)), \delta S \rangle_\Omega 
&= \langle F(u)^\top \cdot \Grad(v), \delta S \rangle_\Omega.
\end{align*}
The identity \eqref{eq:weak:c} is simply the weak form of the boundary condition~\eqref{eq:bc:a}. 
\end{proof}

\subsection{Power balance}

The proof of the following result shows that the power balance of the system is already fully encoded in the weak formulation of the problem, which further also shades light on the role of input and output port variables. 
\begin{proposition} \label{pro:energy}
Let $(v,S,u)$ be a sufficiently regular solution of \eqref{eq:firstorder:a}--\eqref{eq:kinematic-eq} and \eqref{eq:bc:a}--\eqref{eq:bc:b} with $S=S^\top$ and reaction force $\lambda:=F(u)\cdot S\cdot N$ on $\partial\Omega_D$. Then 
\begin{align} \label{eq:energybalance}
\frac{d}{dt}H(\rho v,A:S) = {\langle b, v\rangle}_{\Omega} + {\langle\lambda, v_D\rangle}_{\partial\Omega_D} + {\langle\tau_N, v\rangle}_{\partial\Omega_N}.
\end{align}
\end{proposition}

\begin{proof}
Formally taking the time derivative of the Hamiltonian defined in \eqref{eq:Hamiltonian-elastic} yields
\begin{align*}
\frac{d}{dt} H(\rho v,A:S) 
  = {\langle\rho \dot{v}, v\rangle}_{\Omega} + \langle A:\dot{S}, S\rangle_{\Omega}.
\end{align*}
By using \eqref{eq:weak:a} with $\delta v=v$ and \eqref{eq:weak:b} with $\delta S=S$, we then further see that 
\begin{align*}
\frac{d}{dt} H(\rho v,A:S) 
&= -\langle S,F(u)^\top\cdot\Grad(v)\rangle_{\Omega} 
+ \langle b, v\rangle_{\Omega} 
+ \langle\lambda, v\rangle_{\partial\Omega_D} \\
& \qquad \qquad \qquad 
+ \langle\tau_N, v\rangle_{\partial\Omega_N} 
+ \langle S,F(u)^\top\cdot\Grad(v)\rangle_{\Omega}.
\end{align*}
The first and last term on the right hand side cancel each other, and \eqref{eq:weak:c} allows to replace the third term on the right hand side. This already leads to the claim.
\end{proof}

\begin{remark}
The variational identities \eqref{eq:weak:a}--\eqref{eq:weak:c} do not give a complete characterization of solutions of our problem. Together with the power balance \eqref{eq:energybalance}, they however already encode the underlying PH structure and  reveal the role of the distributed and boundary port variables. 
We recover the ports as indicated already in Remark \ref{remark:phs}.
Again, the kinematic relation \eqref{eq:kinematic-eq} is required to determine the displacement field $u$, which modulates the energy-exchange mechanism. 
\end{remark}

\section{Structure-preserving discretization}
\label{sec:3}

We now illustrate that the variational identities \eqref{eq:weak:a}--\eqref{eq:weak:c} are well-suited for a structure-preserving discretization. An incomplete description (i.e., without considering the kinematic equation) again suffices already to establish a discrete versions of the power balance derived in the previous section. 

\subsection{Spatial discretization}
For the discretization in space, we use a Galerkin approximation of the variational identities \eqref{eq:weak:a}--\eqref{eq:weak:c}. To this end, let us denote by
\begin{align*}
\mathcal{V}_h \subset H^1(\Omega;\mathbb{R}^d), 
\quad 
\mathcal{S}_h\subset L^2(\Omega;\mathbb{R}_{\Sym}^{d\times d}), 
\quad \text{and} \quad  
\Lambda_h\subset L^2(\partial\Omega_D;\mathbb{R}^d)
\end{align*}
appropriate finite dimensional subspaces. 
A particular example will be given in Section~\ref{sec:4} below. 
Under a mild compatibility condition on these spaces and appropriate initial conditions, we will construct an approximation $(v_h,S_h,\lambda_h) : [0,T] \to V_h \times S_h \times \Lambda_h$ for the solution of our problem satisfying the discretized identities
\begin{subequations} \label{eq:approxweak}
 \begin{align} 
	\langle\rho\dot{v}_h,\delta v_h\rangle_{\Omega} &= -\langle S_h,F(u_h)^\top\cdot\Grad(\delta v_h)\rangle_{\Omega} + \langle b,\delta v_h\rangle_{\Omega} \label{eq:approxweak:a}\\
    &\qquad \qquad \qquad + \langle\lambda_h,\delta v_h\rangle_{\partial\Omega_D} + \langle\tau_N,\delta v_h\rangle_{\partial\Omega_N} \notag
\\
	\langle A:\dot{S_h},\delta S_h\rangle_{\Omega} &= \langle F(u_h)^\top\cdot\Grad(v_h),\delta S_h\rangle_{\Omega} 		\label{eq:approxweak:b}
\\
	\langle v_h,\delta\lambda_h\rangle_{\partial\Omega_D} &= \langle v_D,\delta\lambda_h\rangle_{\partial\Omega_D} 		\label{eq:approxweak:c}
\\
\intertext{for all $\delta v_h\in\mathcal{V}_h$, $\delta S_h\in\mathcal{S}_h$, and $\delta \lambda_h\in\Lambda_h$, and for all $t \ge 0$ of relevance.
The discrete displacement field $u_h : [0,T] \to V_h$ will be constructed such that}
\dot u_h &= v_h \label{eq:approxweak:d}
\end{align}
\end{subequations}
which amounts to the discretized kinematic relation. 
Without going into further details, we can already establish the following semi-discrete power-balance.

\begin{proposition}\label{pro:discreteenergy}
Let $(v_h,S_h,\lambda_h,u_h)$ denote a solution of \eqref{eq:approxweak:a}--\eqref{eq:approxweak:d}. Then 
\begin{align} \label{eq:discreteenergybalance}
\frac{d}{dt}H(\rho v_h,A:S_h) &=
\langle b, v_h\rangle_{\Omega} 
+ \langle\lambda_h, v_D\rangle_{\partial\Omega_D} 
+ \langle\tau_{N}, v_h\rangle_{\partial\Omega_N}.
\end{align}
\end{proposition}
The proof of this result follows with the very same arguments as that of Proposition~\ref{pro:energy} and is therefore omitted. 
Let us note that, like on the continuous level, the variational identities \eqref{eq:approxweak:a}--\eqref{eq:approxweak:c} already encode the power balance of the system, but the kinematic relation \eqref{eq:approxweak:d} is additionally required to determine the modulating discrete displacement field which encodes the details of the energy-exchange mechanism.

\subsection{Differential-algebraic form}

By choosing some basis for the approximation spaces $V_h$, $S_h$, $\Lambda_h$, the variational identities \eqref{eq:approxweak:a}--\eqref{eq:approxweak:c} can be turned into a system of differential-algebraic equations
\begin{align} \label{eq:statespace}
\begin{bmatrix}
  \ttM_\ttv & 0 & 0\\
  0 & \ttM_\ttS & 0 \\
	 0 & 0 & 0
\end{bmatrix}
\begin{bmatrix}
\dot{\ttv} \\ \dot{\ttS} \\ \dot{\ttlam}
\end{bmatrix} 
&=
\begin{bmatrix}
0 & -\ttD(\ttu) & \ttB \\
\ttD(\ttu)^\top & 0 & 0\\
-\ttB^\top & 0 & 0
\end{bmatrix}
\begin{bmatrix}
\ttv \\ \ttS \\ \ttlam
\end{bmatrix} 
+
\begin{bmatrix}
\ttG_\ttb & \ttG_\tttau & 0 \\ 
0 & 0 & 0 \\ 
0 & 0 & \ttG_\ttv
\end{bmatrix}
\begin{bmatrix}
\ttb \\ \tttau_\ttN \\ \ttv_\ttD
\end{bmatrix},
\end{align}
where $\ttv$, $\ttS$, $\ttlam$, and so on, are the vector representations of the corresponding discretized functions; see \cite{Bonet2008,Wriggers2001} for details. 
Let us note that systems of similar structure were also obtained in~\cite{Brugnoli2021a,thoma2022} for beams and strings.
The discretized kinematic constraint finally has the simple form 
\begin{align} \label{eq:statespace:kinematics}
    \dot{\ttu} = \ttv.
\end{align}
Since no time derivative of the Lagrange multiplier $\ttlam$ appears and the last equation on the right hand side of \eqref{eq:statespace} is independent of $\ttlam$, the above system amounts to a differential-algebraic equation of index $\nu \ge 2$; see e.g. \cite{KunkelMehrmann2006}. 
The variables $\ttv$, $\ttS$, and $\ttu$ are of differential nature, while $\ttlam$ is the algebraic variable.  
\begin{lemma}
Assume that $\ttB^\top \cdot \ttB$ is regular.
Then the above system is a regular differential-algebraic equation with index $\nu=2$. 
\end{lemma}
\begin{proof}
After reordering, the system can be seen to be of Hessenberg form 
\begin{align} \label{eq:hessenberg}
\ttM \dot \tty = \ttf(\tty,\ttz), \qquad 
0 = \ttg(\tty)
\end{align}
with regular matrix $\ttM$ in front of the time derivatives, and variables $\tty=(\ttv,\ttS,\ttu)$ and $\ttz=\ttlam$.
By differentiating the second equation, one obtains the \emph{hidden constraint}
\begin{align} \label{eq:hidden}
0 &= \ttg_\tty(\tty) \cdot \dot \tty 
= \ttg_\tty(\tty) \cdot \ttM^{-1} \cdot \ttf(\tty,\ttz) =: \tth(\tty,\ttz).
\end{align}
This equation allows to uniquely determine $\ttz$ as a function of $\tty$, if  
\begin{align} \label{eq:index2}
\partial_\ttz \tth(\tty,\ttz) = \ttB^\top \ttM^{-1} \ttB \qquad \text{is regular}. 
\end{align}
Since $\ttM$ is regular, this condition is equivalent to $\ttB^\top \cdot \ttB$ being regular.
Hence a differential equation for $\dot z$ can be obtained by differentiating the hidden constraint and, as a consequence, the index of the system is $\nu=2$; we refer to~\cite{KunkelMehrmann2006} for details.
\end{proof}

\begin{remark}
The regularity condition for $\ttB^\top \cdot \ttB$ is equivalent to assuming that $\ttB$ is injective, which is required to uniquely determine the Lagrange multiplier $\ttlam$. 
On the level of the Galerkin approximation, this means that 
\begin{align}
\langle \lambda_h, \delta v_h\rangle_{\partial\Omega_D}=0 \quad \forall \delta v_h 
\qquad \text{implies that} \qquad \lambda_h=0.
\end{align}
This amounts to a compatibility condition for the approximation spaces $V_h$, $\Lambda_h$, which can be satisfied, e.g., by choosing $\Lambda_h=V_h|_{\partial\Omega_D}$; see Section~\ref{sec:4} below.
Uniqueness of a solution for appropriate initial values is then guaranteed by regularity of the system \eqref{eq:hessenberg}. 
In order to guarantee existence of a solution $(y,z)=(\ttv,\ttS,\ttlam,\ttu)$, the initial values have to satisfy the compatibility conditions 
\begin{align} \label{eq:compatibility}
\ttg(\tty(0))=0 \qquad \text{and} \qquad \tth(\tty(0),\ttz(0)) = 0.
\end{align}
These conditions can be translated directly to conditions for $\ttv(0)$, $\ttS(0)$, $\ttlam(0)$, $\ttu(0)$ or the corresponding discrete functions. 
\end{remark}

\subsection{Time discretization}
We now demonstrate that the particular geometric structure and the corresponding power-balance are also preserved by appropriate time discretization. 
Let $\Dtau>0$ be a fixed step size and $t^n = n \Dtau$, $n \ge 0$ be the discrete time steps. 
We define
\begin{align*}
\dtau a^{n+1/2} = \frac{1}{\tau}(a^{n+1} - a^n) 
\qquad \text{and} \qquad 
a^{n+1/2} = \frac{1}{2}(a^{n+1} + a^n)
\end{align*}
to denote the difference and average of a sequence of numbers defined over the time grid. 
Applying the mid-point rule to \eqref{eq:approxweak:a}--\eqref{eq:approxweak:c} leads to 
\begin{subequations} \label{eq:discrete-time:weakform}
 \begin{align} 
	\langle\rho\Dtau v_h^{n+1/2},\delta v_h\rangle_{\Omega} &= -\langle S_h^{n+1/2},F(\bar u_h^{n+1/2})^\top\cdot\Grad(\delta v_h)\rangle_{\Omega}  \label{eq:discrete-time:weakform:a}\\
    &\hspace*{-2em} + \langle b^{n+1/2},\delta v_h\rangle_{\Omega} + \langle\lambda_h^{n+1/2},\delta v_h\rangle_{\partial\Omega_D} + \langle\tau_N^{n+1/2},\delta v_h\rangle_{\partial\Omega_N} \notag
\\
	\langle A: \Dtau S_h^{n+1/2},\delta S_h\rangle_{\Omega} &= \langle F(\bar u_h^{n+1/2})^\top\cdot\Grad(v_h^{n+1/1}),\delta S_h\rangle_{\Omega} 		\label{eq:discrete-time:weakform:b}
\\
	\langle v_h^{n+1/2},\delta\lambda_h\rangle_{\partial\Omega_D} &= \langle v_D^{n+1/2},\delta\lambda_h\rangle_{\partial\Omega_D} 		\label{eq:discrete-time:weakform:c}
\\
\intertext{for all $\delta v_h\in\mathcal{V}_h$, $\delta S_h\in\mathcal{S}_h$, and $\delta \lambda_h\in\Lambda_h$, and for all time steps $n \ge 0$ of relevance.
The discrete displacement field is updated according to}
\dtau \bar u_h^{n} &= v_h^{n-1/2}. \label{eq:discrete-time:weakform:d}
\end{align}
\end{subequations}
As could be seen on the continuous and the semi-discrete level, the power-balance of the system is tightly coupled to the particular form of the variational identities. This remains true for the fully-discrete setting and leads to the following result.
\begin{proposition} \label{pro:discrete-time}
Let $(v_h^n,S_h^n,\lambda_h^n,\bar u_h^{n+1/2})$, $n \ge 0$ be a solution of \eqref{eq:discrete-time:weakform:a}--\eqref{eq:discrete-time:weakform:d}. 
Then 
\begin{align} \label{eq:discrete-time:energybalance}
&\dtau H(\rho v_h^{n+1/2},A:S_h^{n+1/2}) \\
&\qquad \qquad = 
\langle b^{n+1/2}, v_h^{n+1/2}\rangle_{\Omega} 
+ \langle\lambda_h^{n+1/2}, v_D^{n+1/2}\rangle_{\partial\Omega_D} 
+ \langle\tau_{N}^{n+1/2}, v_h^{n+1/2}\rangle_{\partial\Omega_N}. \notag
\end{align}
\end{proposition}
\begin{proof}
It is not difficult to check that 
$\frac{1}{2\Dtau} (a^2 - b^2) 
= \left(\frac{a-b}{\Dtau} \right) \cdot \left( \frac{a+b}{2}\right)$.
This yields 
\begin{align*}
\dtau H(\rho v_h^{n+1/2},A:S_h^{n+1/2}) 
= \langle \rho \dtau v_h^{n+1/2}, v_h^{n+1/2}\rangle_\Omega + \langle A : \dtau S_h^{n+1/2},S_h^{n+1/2}\rangle_{\Omega}.
\end{align*}
The power-balance then follows immediately from the variational identities \eqref{eq:discrete-time:weakform:a}--\eqref{eq:discrete-time:weakform:c} by  
using $\delta v_h = v_h^{n+1/2}$, $\delta S_h = S_h^{n+1/2}$, and $\delta \lambda_h = \lambda_h^{n+1/2}$ as the test functions.
\end{proof}

\begin{remark}
Let us note that the kinematic relation~\eqref{eq:discrete-time:weakform:d} is not required in the proof of the power-balance. Hence there is some freedom to define the update formula for the discrete displacement field. 
The above choice means that $\bar u_h^{n+1/2} = \bar u_h^{n-1/2} + \Dtau v_h^n$ can be computed before solving \eqref{eq:discrete-time:weakform:a}--\eqref{eq:discrete-time:weakform:c} for $v_h^{n+1}$, $S_h^{n+1}$, and $\lambda_h^{n+1}$.
As a consequence, only a single linear system has to be solved in every time iteration.
The iteration \eqref{eq:discrete-time:weakform:d} requires a specific treatment for the first step, e.g. $\bar u_h^{1/2} = u_h^0 + \frac{\Dtau}{2} v_h^{0}$.
The data $b^{n+1/2}$, $\tau_N^{n+1/2}$, $v_D^{n+1/2}$ can also be chosen in different ways, e.g., as the time averages on the respective time intervals or the evaluations at the midpoints $t^{n+1/2}$. 
\end{remark}

\section{Numerical tests}
\label{sec:4}

In order to illustrate the theoretical results derived in the previous sections, we now present some numerical results for a typical model problem. In addition, we discuss some details of the implementation that were not required for the analysis before.  

\subsection{Model problem}

\begin{figure}[htbp]
	\centering
	\def\svgwidth{10cm}
	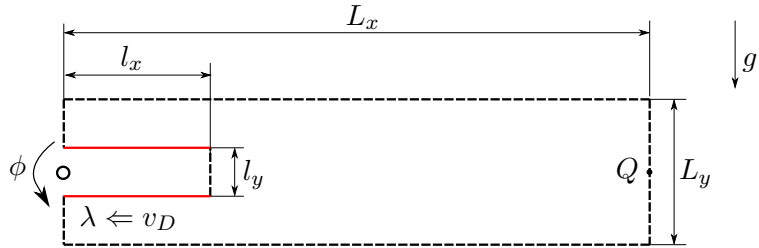
	\caption{Schematic sketch of the geometry. Red lines mark the Dirichlet boundary $\partial\Omega_D$ for which the displacement velocity $v_D$ is prescribed. The Lagrange multiplier corresponds to the reaction forces.}
	\label{fig:experiment}
\end{figure}

We consider a compliant mechanism as part of a soft robotic arm. A rubber-type body is attached to a pivot, whose rotation angle can be controlled. We consider a planar motion under gravity. The angle of the rigid pivot is prescribed as
\begin{align}
	\phi(t) = \begin{cases}
		0,\quad &t\leq 0, \\
		\frac{\pi}{4}\left(10(\frac{t}{T_c})^3 - 15(\frac{t}{T_c})^4 + 6(\frac{t}{T_c})^5\right),\quad &0<t\leq T_c, \\
		\frac{\pi}{4},\quad &t>T_c. \\
	\end{cases}
\end{align}
The corresponding velocity is imposed as Dirichlet boundary condition \eqref{eq:bc:a} to the contact areas $\partial\Omega_D$, depicted in red in Figure~\ref{fig:experiment}. 
The initial conditions at time $t=0$ are given by 
\begin{align}
	v(0)=0, \quad S(0)=S_0, \quad u(0)=u_0, \quad \lambda(0)=\lambda_0,
\end{align}
and the initial data $S_0$, $u_0$, $\lambda_0$ are obtained by solving the corresponding static elasticity problem. 
One can verify that these data satisfy the compatibility conditions \eqref{eq:compatibility}.
The system is modelled in plane strain and its further parameters are listed in Table \ref{tab:parameter}.
\begin{table}[htbp]
\caption{Parameters of the numerical experiment}
	\label{tab:parameter}
        \centering \medskip
	{\small
        \begin{tabular}{ l l l l l l l l l }
		\hline
		$L_x$ & 
		$L_y$ &
		$l_x$ & 
		$l_y$ & 
		$\rho$ & 
		$E$ & 
		$\nu$ & 
		$g$ & 
		$T_c$  \\
        \hline
        $60\;\text{cm}$ & $15\;\text{cm}$ & $15\;\text{cm}$ & $5\;\text{cm}$ & $960\;\text{kg/m}^3$ & $6\cdot10^6\:\text{N/m}^2$ & $0.49$ & $9.81\;\text{m/s}^2$ & $0.5\;\text{s}$ \\
		\hline
	\end{tabular}
    }
\end{table}

\subsection{Details on the discretization}
A mixed finite element method is used for the spatial discretization. 
The domain is partitioned into a uniform triangular mesh $\T_h$ and the Galerkin approximation spaces are defined as 
\begin{align*}
V_h = [P_1(\T_h) \cap H^1(\Omega)]^2, \qquad S_h = P_0(\T_h;\RR^{2 \times 2}_{\Sym}), \qquad 
\Lambda_h = V_h|_{\partial\Omega_D}.
\end{align*}
The mesh size for our tests is chosen as $h\approx9\;\text{mm}$. The time step is set to $\Dtau = 2\;\text{ms}$ and the simulations are run until final time $T=1\;\text{s}$. 
All computations were realized in the finite element framework FEniCS
~\cite{LoggMardalEtAl2012,Kuchta2020}.

\subsection{Computational results}
\begin{figure}[htbp]
	\centering
	\def\svgwidth{10cm}
\begingroup%
  \makeatletter%
  \providecommand\color[2][]{%
    \errmessage{(Inkscape) Color is used for the text in Inkscape, but the package 'color.sty' is not loaded}%
    \renewcommand\color[2][]{}%
  }%
  \providecommand\transparent[1]{%
    \errmessage{(Inkscape) Transparency is used (non-zero) for the text in Inkscape, but the package 'transparent.sty' is not loaded}%
    \renewcommand\transparent[1]{}%
  }%
  \providecommand\rotatebox[2]{#2}%
  \newcommand*\fsize{\dimexpr\f@size pt\relax}%
  \newcommand*\lineheight[1]{\fontsize{\fsize}{#1\fsize}\selectfont}%
  \ifx\svgwidth\undefined%
    \setlength{\unitlength}{1642.5bp}%
    \ifx\svgscale\undefined%
      \relax%
    \else%
      \setlength{\unitlength}{\unitlength * \real{\svgscale}}%
    \fi%
  \else%
    \setlength{\unitlength}{\svgwidth}%
  \fi%
  \global\let\svgwidth\undefined%
  \global\let\svgscale\undefined%
  \makeatother%
  \begin{picture}(1,0.52785389)%
    \lineheight{1}%
    \setlength\tabcolsep{0pt}%
    \put(0,0){\includegraphics[width=\unitlength,page=1]{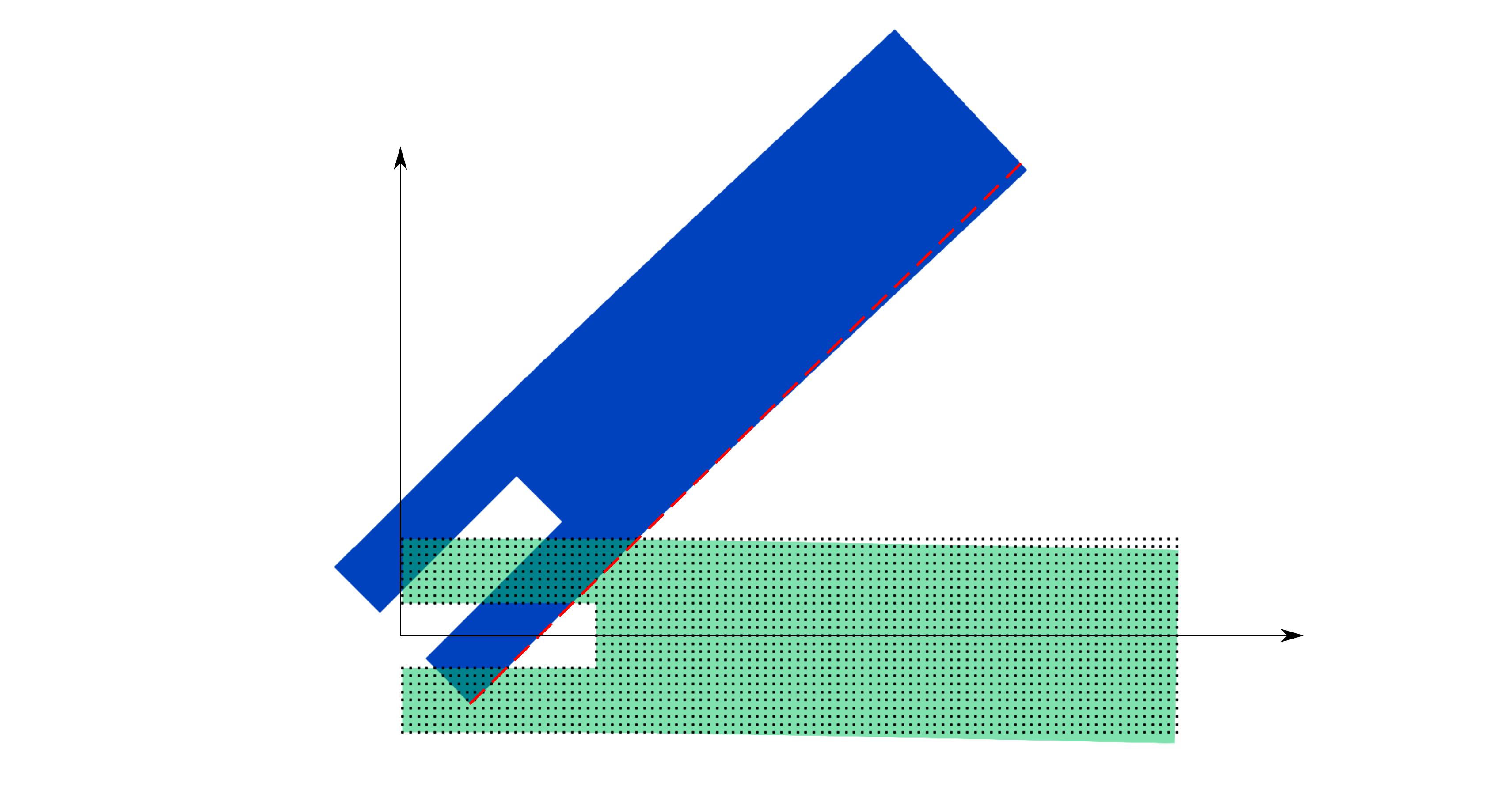}}%
    \put(0.87297398,0.10294761){\color[rgb]{0,0,0}\makebox(0,0)[lt]{\lineheight{1.25}\smash{\begin{tabular}[t]{l}$X_1$\end{tabular}}}}%
    \put(0.78497107,0.03237925){\color[rgb]{0,0,0}\makebox(0,0)[lt]{\lineheight{1.25}\smash{\begin{tabular}[t]{l}$t=0\;\text{s}$\end{tabular}}}}%
    \put(0.69780821,0.4142742){\color[rgb]{0,0,0}\makebox(0,0)[lt]{\lineheight{1.25}\smash{\begin{tabular}[t]{l}$t=0.722\;\text{s}$\end{tabular}}}}%
    \put(0.21751922,0.43972515){\color[rgb]{0,0,0}\rotatebox{2.3537804}{\makebox(0,0)[lt]{\lineheight{1.25}\smash{\begin{tabular}[t]{l}$X_2$\end{tabular}}}}}%
    \put(0,0){\includegraphics[width=\unitlength,page=2]{paraview.pdf}}%
  \end{picture}%
\endgroup%

	\caption{Snapshots at $t=0\;\text{s}$ (green) and $t=0.722\;\text{s}$ (blue), dots represent the reference configuration $\Omega$, and the red dashed lines demonstrate the rigid body configuration.  Images generated with ParaView~\cite{Ayachit2015}.}
	\label{fig:snapshots}
\end{figure}

The motion of the robot arm can be approximately described as that of a rigid body. 
In Figure~\ref{fig:snapshots}, we show two snapshots of the motion of the rigid and the elastic body.
As expected, the absolute deviations in the displacement are rather small, which to some extent justifies the use of a rigid body model as a surrogate in simulations. 

\begin{figure}[htbp]
	\centering
	\includegraphics[width=0.7\textwidth]{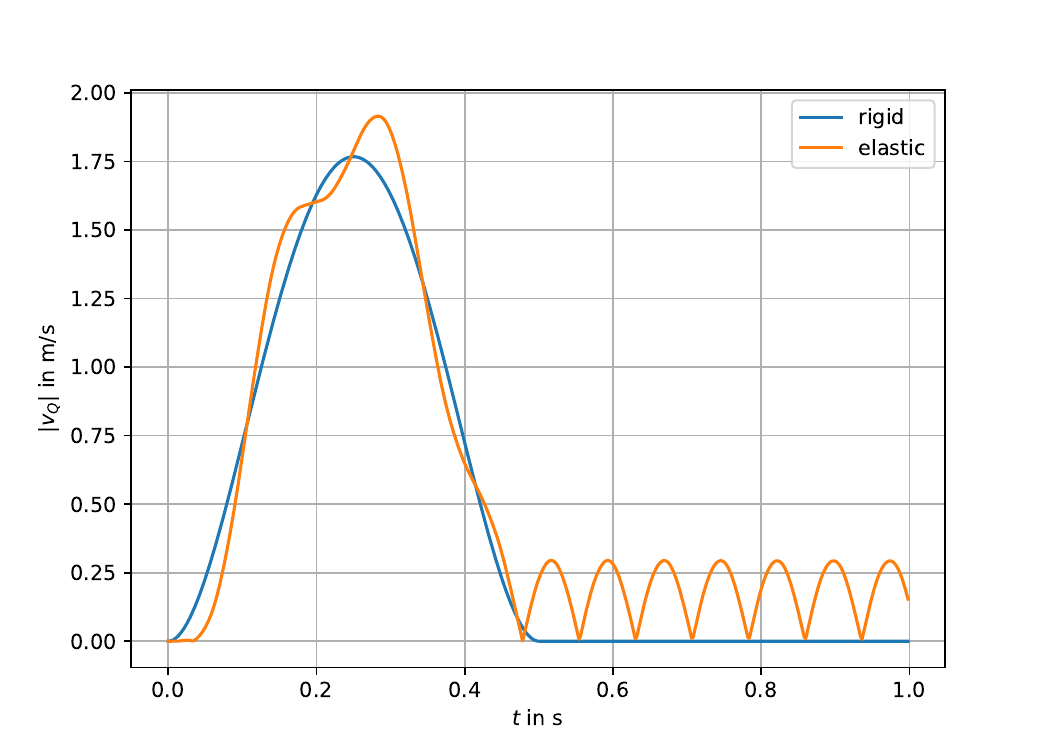}
	\caption{Absolut velocities $|v|$ at the point $Q=(L_x,0)$ for a rigid body motion (blue) compared to the true elastic motion (orange). Elastic vibrations pertain after the motion of the pivot has been stopped.}
	\label{fig:velocity}
\end{figure}

In Figure~\ref{fig:velocity}, we illustrate in more detail the differences in velocity between the simplified rigid body and the elastic model considered in this paper. 
Note that the time variation  of the angle $\phi(t)$ leads to elastic deformations causing vibrations of the body. Due to lack of a damping mechanism, the vibrations do not fade away when stopping the motion of the pivot at time $t \ge 0.5\,\text{s}$. 
In the flexible  case, a more sophisticated trajectory planning based on the elastic model is needed for the reduction of such structure-induced oscillations, complemented by feedback control, if necessary; see e.g. \cite{Wang2021}. 

\begin{figure}[htbp]
	\centering
	\includegraphics[width=0.7\textwidth]{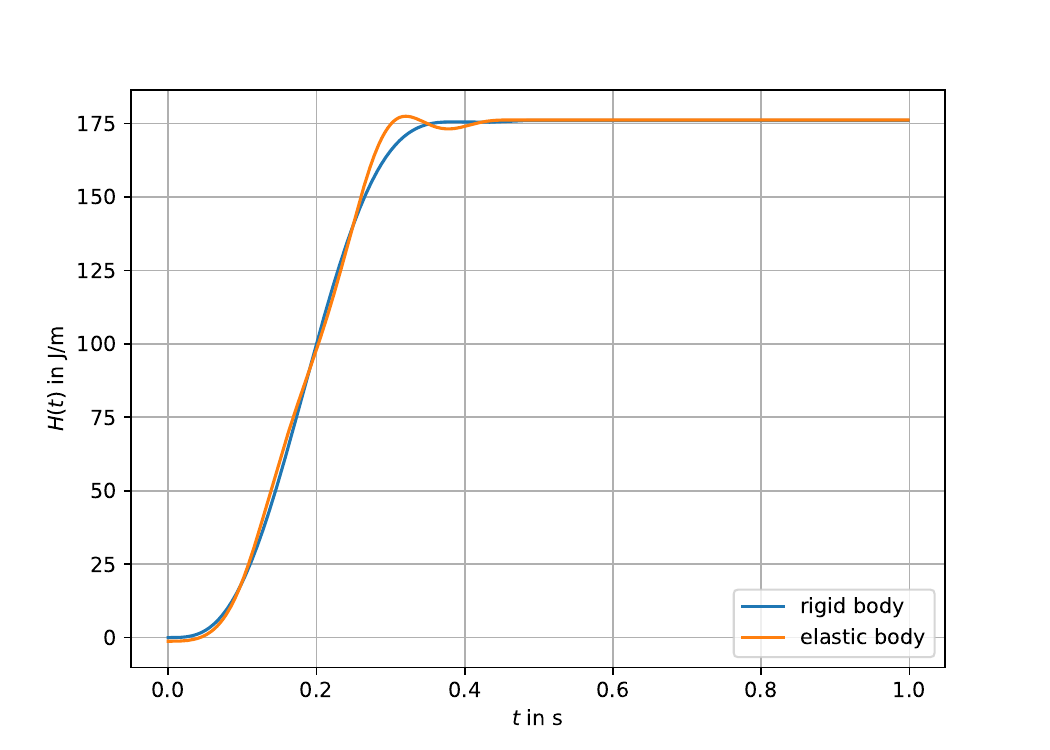}
	\caption{Approximation of the total energy $H_{tot}$ obtained for the rigid body model and the structure-preserving fully discrete approximation of the elasticity system proposed in this paper.}
	\label{fig:energy}
\end{figure}

As a last step, we would like to illustrate the exact preservation of energy due to the discrete power balance derived in the previous section. 
To do so, we consider the total energy of the system 
\begin{align}
H_{tot}(v,S,u) := H(v,S) + H_{gra}(u),
\end{align}
consisting of the internal energy $H(v,X)=\frac{1}{2}\langle \rho v,v \rangle_\Omega + \frac{1}{2}\langle A : S, S\rangle_\Omega$ and the potential energy $H_{gra}(u)=-\langle b,u\rangle_\Omega$ due to the graviational forces. 
The approximation of the total energy in the fully discrete setting is then given by 
\begin{align}
H_{h,tot}^n = H(v_h^n,S_h^n) + H_{gra}(\hat u_h^n)
\end{align}
with $\hat u_h^n = u_h^{n-1/2} + \frac{\Dtau}{2} v_h^{n} = u_h^{n+1/2}-\frac{\Dtau}{2} v_h^{n+1}$.
With this approximation, the discrete power balance stated in Proposition~\ref{pro:discrete-time} can be rephrased as 
\begin{align}
\dtau H_{h,tot}^{n+1/2} =  \langle\lambda_h^{n+1/2}, v_D^{n+1/2}\rangle_{\partial\Omega_D} 
+ \langle\tau_{N}^{n+1/2}, v_h^{n+1/2}\rangle_{\partial\Omega_N}.
\end{align}
In our test example, we have $v_D=0$ and $\tau_N=0$ for all $t \ge 0.5\;\text{s}$. 
Hence the total energy of the discrete system should be conserved after this time instant.
In Figure~\ref{fig:energy}, we display the corresponding results.
As predicted by our theoretical results, the total energy of the fully discrete systems stays exactly constant after time $t\geq0.5\;\text{s}$.

\section{Discussion}
\label{sec:5}

The motion of a geometrically nonlinear elastic continuum was considered by means of a velocity-stress formulation. 
The governing equations have the form of a PH system in which the skew-symmetric energy-transfer operator is modulated by the displacement field. 
Based on a weak formulation of the problem, in which the Dirichlet boundary conditions are incorporated by Lagrange multipliers, we derived a power balance and discussed the structure-preserving discretization by mixed finite element methods in space and an implicit midpoint rule in time. 
A different time integration of the kinematic relation for the displacement field led to a linear implicit time-stepping scheme. 
The power balance of the system could be proven to be preserved on the semi-discrete and fully-discrete level. 
The main theoretical results were illustrated by numerical tests for a typical application. 

In summary, we obtained a structure-preserving discretization for geometrically nonlinear elastodynamics, which can be realized at the same computational effort as in the linear case. 
A key step to achieve this was to utilize the PH structure of the velocity-stress formulation, while treating the displacement field as an additional passive kinematic variable. 
The restriction to St. Venant–Kirchhoff materials used in this paper is suitable for application with large deformation but small strains. The extension to more complex nonlinear energies including multiphysical effects and damping mechanisms is topic of furture research.

\bibliographystyle{tfq}
\bibliography{mybib}

\end{document}